\documentclass[conference]{IEEEtran}
\IEEEoverridecommandlockouts

\usepackage{cite}
\usepackage{amsmath,amssymb,amsfonts}
\usepackage{graphicx}
\usepackage{textcomp}
\usepackage{xcolor}
\usepackage{mathrsfs}
\usepackage{cite}
\usepackage{array}
\usepackage{tabularx}
\usepackage[table]{xcolor}
\usepackage{multicol, multirow}
\usepackage{color}
\usepackage{mathtools}
\usepackage{comment}
\usepackage{stmaryrd}
\usepackage{algpseudocode}
\usepackage{subcaption}
\usepackage{tikz,pgf}
\usepackage{pgfplots}
\usepackage{verbatim}
\usetikzlibrary{calc,positioning,mindmap,trees,decorations.pathreplacing}
\usepackage{graphicx}
\usepackage{bbm}
\usepackage[utf8]{inputenc}
\usepackage{hyperref}
\usepackage{bm}

\newcommand{\argmax}{\mathop{\rm arg~max}\limits}

\DeclarePairedDelimiter\abs{\lvert}{\rvert}
\DeclarePairedDelimiter\norm{\lVert}{\rVert}
\DeclareMathOperator{\diag}{diag}

\usepackage[colorinlistoftodos,prependcaption,backgroundcolor=black!5!white,bordercolor=red]{todonotes}

\def\BibTeX{{\rm B\kern-.05em{\sc i\kern-.025em b}\kern-.08em
    T\kern-.1667em\lower.7ex\hbox{E}\kern-.125emX}}
\begin{document}
\pagestyle{plain}

\title{Enabling Drone Detection with SWARM Repeater-Assisted MIMO ISAC\\

}

\author{\IEEEauthorblockN{Palatip Jopanya and Diana P. M. Osorio} \IEEEauthorblockA{ {Department of Electrical Engineering,} {Link\"oping University}, Sweden }  \IEEEauthorblockA{E-mail: \{palatip.jopanya, diana.moya.osorio\}@liu.se} }

\maketitle

\begin{abstract}
As definitions about new architectural aspects, use cases, and standards for integrated sensing and communication (ISAC) continue to appear, cellular systems based on massive multiple-input multiple-output (MIMO) antenna technology are also experiencing a parallel evolution through the integration of novel network components. This evolution should support emerging ISAC use cases and services. In particular, this paper explores a recent vision for cost-efficient cellular network densification through the deployment of swarms of repeaters. Leveraging their ability to retransmit signals instantaneously, we investigate how these repeaters can enhance radar sensing capabilities for drone detection in a swarm repeater-assisted MIMO ISAC system. 
\end{abstract}

\begin{IEEEkeywords}
drone detection, integrated sensing and communications, MIMO ISAC, swarm repeaters.
\end{IEEEkeywords}

\section{Introduction}
Multiple-input multiple-output (MIMO) technology has been at the core of mobile networks, specially with massive MIMO in the fifth-generation (5G), delivering gains in spatial diversity, multiplexing, and beamforming. Toward the sixth-generation (6G), the concept of distributed MIMO is evolving, particularly under the promise to serve as enabling technology for networked integrated sensing and communication (ISAC)~\cite{11021487}. However, blind spots may occur, leading to fewer MIMO data streams for users, and weak links for target sensing. 

In this way, new network components have emerged that promise to satisfy these requirements, such as the network-controlled repeater (NCR). More specifically, its evolution into a network component operating as a full-duplex relay, instantaneously amplifying and re-transmitting signals with only a few hundred nanoseconds of delay \cite{10908557}, can appear as part of future scalable and adaptive networks. 
It is expected that by deploying this new component at large scale, in a swarm repeater-assisted cellular massive MIMO, promising performance enhancements can be achieved (approaching those expected from distributed MIMO systems) while maintaining low cost and low power consumption, as envisioned in~\cite{10908557}. The key advantage of repeater-assisted cellular networks is that they act as scatterers on the environment, thus improving the channel rank and overcoming coverage holes. With a careful selected amplification gain in a repeater-assisted cellular network, the system remains stable and avoids positive feedback caused by repeater interactions \cite{11322715}. 

Among the potential use cases for ISAC, drone intrusion detection is highlighted as a highly demanded and timely application, thus introduced in the technical report on ISAC feasibility studies~\cite{TR}. Research on this field is gaining momentum, for instance, unauthorized drone detection is investigated in~\cite{detunauth} for a cell-free MIMO ISAC system. The work introduced age of sensing and sensing coverage as performance metrics, and proposed joint blocklength–power optimization. In~\cite{jopaUtiliSSB}, a scheme for passive sensing of a drone is proposed based on the Synchronization Signal Block (SSB) sweeping beams in a bistatic setting, where Cr\'{a}mer-Rao bounds (CRB) for range and velocity were derived. In~\cite{7735118}, the author proposes a passive radar system using a uniform circular array to detect and track UAVs by capturing OFDM echoes emitted from nearby base stations. In \cite{9716078}, the author introduces and demonstrates the concept of using cellular base stations for imaging and UAV detection, functioning similarly to inverse synthetic-aperture radar.

Considering the integration of two technologies, ISAC and swarm repeater-assisted MIMO networks, the study in~\cite{11226628} addresses the key question on what would be the impact on sensing performance with the introduction of repeaters. To that, the work derived the CRB for range estimation under repeater interference and proposed a joint optimization of repeater amplification and access point (AP) precoder. Besides this initial work, several questions remain unexplored for this type of networks. 

To address this gap, herein we investigate how to exploit swarm repeater-assisted MIMO ISAC networks to support the sensing functionality. Particularly, we consider multiple single-antenna amplify-and-forward repeaters, each with independently configured amplification gain. An AP simultaneously serves a user in a dense urban environment and detects a drone in a weak coverage area. We consider a monostatic MIMO ISAC setup with a line-of-sight (LOS) channel between the AP and the target. With the presence of swarm repeaters, the system optimizes the gain of each repeater to maximize the sensing signal-to-interference-plus-noise ratio (SINR) for drone detection while guaranteeing the user's SINR.



\begin{figure}[htbp]
    \centering
        \includegraphics[width=3in]{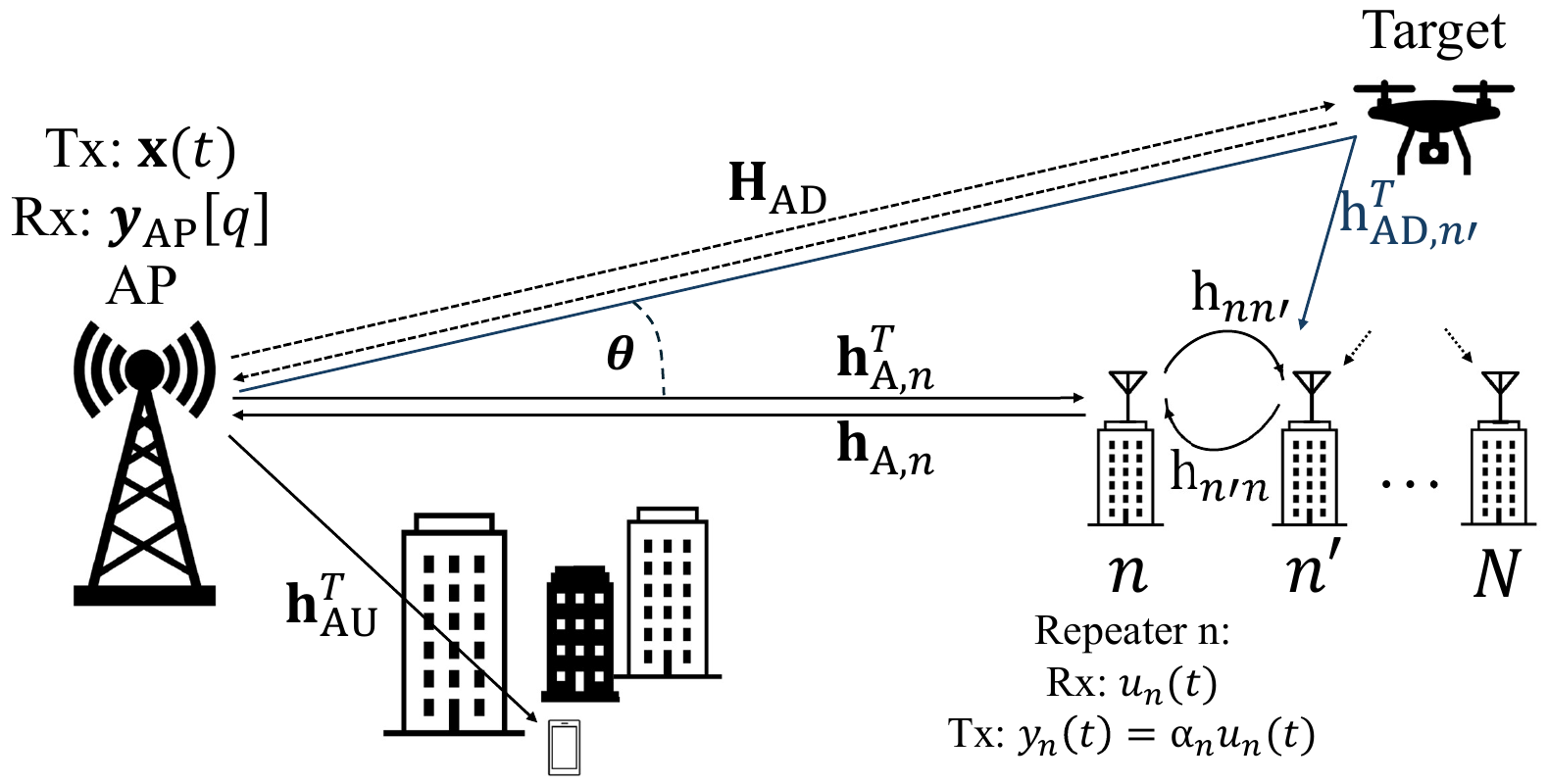}
        \caption{System model.}
    \label{fig:system_model}
\end{figure} 
\section{System model}
We consider a monostatic MIMO ISAC AP equipped with $M$ antennas arranged in a uniform linear array (ULA). The AP simultaneously serves a user equipment (UE) and detects a drone located in a weak coverage region. In addition, $N$ repeaters are deployed and arranged in a line with uniform spacing $d$ meters at the same altitude as the AP. The $n$th repeater is located $l_{\text{A},n}$ meters from AP. The repeater index $n$ is ordered in ascending order, from the closest repeater to the farthest repeater from the AP. At repeater $n$, the received signal is retransmitted with gain $\alpha_n ;\ \alpha_n \in \mathbb{C}$. The angle $\theta$ is the true angle of the target drone, while $\hat{\theta}$ is an angle of beamforming vector in the DFT codebook that gives the highest echoes under single target assumption. The channel from the AP to the UE is modeled as Rayleigh fading, while the channels from the AP to the drone, from the drone to the repeaters, between repeaters, and from repeaters to AP are LOS. The system model can be illustrated in Figure \ref{fig:system_model}. Note that the links between the UE and the repeaters are weak and are neglected for simplicity.

\subsection{Channel models}
The two-link LOS channel of a monostatic MIMO ISAC system, AP-drone-AP, is 
\begin{align*}
    \mathbf{H}_{\text{AD}}= r \sqrt{\beta_\text{AD}}e^{-j2\pi f_c \tau_\text{AD}}\mathbf{a}({\theta})\mathbf{a}^{\top}({\theta}),
\end{align*}
where $\beta_\text{AD}={\lambda^2}/{((4\pi)^3l_\text{AD}^4)}$ is the large-scale fading coefficient, $\tau_{\text{AD}}$ is the round trip propagation delay, the steering vector $\mathbf{a}(\theta)=[1, e^{j\pi \cos (\theta)},\dots,e^{j(M-1)\pi \cos (\theta)}]^{\top}$, $\lambda$ is the carrier wavelength, $r\sim \mathcal{CN}(0,\sigma^2_{\text{rcs}})$ is the radar cross section (RCS), and $l_\text{AD}$ is the distance between the AP and the drone. The LOS channel from the AP to the repeater $n$ is $\mathbf{h}_{\text{A},n}=\sqrt{\beta_{\text{A},n}}e^{-j2\pi f_c \tau_{\text{A},n}}\mathbf{a}(0),$
where $\beta_{\text{A},n}={\lambda^2}/{(4\pi l_{\text{A},n})^2}$, $l_{\text{A},n}$ is the AP-repeater-$n$ distance, and $\tau_{\text{A},n}$ is the propagation delay. The two-link LOS channel from AP–drone–repeater $n$ is modeled as $\mathbf{h}_{\text{AD},n} = r\sqrt{\beta_{\text{AD},n}}e^{-j2\pi f_c\tau_{\text{AD},n}}\mathbf{a}(\theta)$, where $\beta_{\text{AD},n}={\lambda^2}/{((4\pi)^3 (l_\text{AD}l_{\text{D},n})^2)}$, with the distance from drone to repeater $n$, $l_{\text{D},n}$. The LOS channel between repeater $n$ and $n'$ is $h_{nn'}= \sqrt{\beta_{nn'}}e^{-j2\pi f_c \tau_{nn'}}$, where $\tau_{nn'}$ is a propagation delay between repeater $n$ and $n'$, $\beta_{nn'}={\lambda^2}/{(4\pi l_{nn'})^2}$, and $l_{nn'}$ is the distance between repeater $n$ and $n'$. The inter-repeater channel matrix $\mathbf{H}_{\text{RR}} $$\in$$ \mathbb{C}^{N\times N}$ is defined as $[\mathbf{H}_{\text{RR}}]_{n,n'}=h_{nn'}$. The channel from AP to UE is modeled as $\mathbf{h}_{\text{AU}}$$\sim$$ \mathcal{CN}(0,\beta_{\text{AU}}\mathbf{I}_M)$, where $\beta_{\text{AU}}={\lambda^2}/{(4\pi l_{\text{AU}})^2}$, and $l_{\text{AU}}$ is the distance between AP and UE. Note that the channel between repeater $n$ and $n'$ is reciprocal, $h_{nn'}=h_{n'n} , \forall n,n'$. The self-interference at each repeater is assumed to be removed, i.e., $h_{nn}=0 , \forall n$. 
\subsection{Signal models}
\subsubsection{Transmitted signal} The transmitted signal at time $t$, $\mathbf{x}(t) \in \mathbb{C}^M$, is 
\begin{equation}
    \label{eqn:tx_AP}
    \mathbf{x}(t) = \Bigl( \sqrt{\rho_\text{s}} \mathbf{w}_\text{s} s_\text{s}  + \sqrt{\rho_\text{c}} \mathbf{w}_\text{c} s_\text{c} \Bigl) e^{j2\pi f_c t},
\end{equation}
where $\rho_\text{s}$ is the sensing power, $\rho_\text{c}$ is the communication power, $\mathbf{w}_\text{s} $$\in$$ \mathbb{C}^M$ is a sensing precoder, $\mathbf{w}_\text{c} $$\in$$ \mathbb{C}^M$ is a communication precoder, $s_\text{s}$ is the sensing symbol, $s_\text{c}$ is the communication symbol, and $f_c$ is the carrier frequency. $s_\text{s}$ and $s_\text{c}$ have zero mean, uncorrelated, and unit power $\mathbb{E}[\abs{s_\text{s}}^2]=\mathbb{E}[\abs{s_\text{c}}^2]=1$.

\subsubsection{Received and retransmitted signal at the repeaters}
The signal received at repeater $n$ consists of three components: the downlink signal transmitted from AP, interaction from all the other repeaters, and echoes from the drone. The signal received at repeater $n$, $u_n \in \mathbb{C}$, is 
\begin{equation}
    \begin{split}
    \label{received_repeater_n}
    u_n(t) &= \underbrace{\mathbf{h}_{\text{A},n}^{\top} \mathbf{x}(t)}_\text{Downlink from AP} + \underbrace{\sum_{n'\neq n}h_{nn'}y_{n'}(t)}_\text{Repeaters interaction} \\
   &+ \underbrace{\mathbf{h}^{\top}_{\text{AD},n}\mathbf{x}(t)}_\text{Echoes (drone)} + \underbrace{n_\text{R}(t)}_\text{Noise (repeater)},
    \end{split}
\end{equation}
where $y_{n'}(t)$ is the signal retransmitted by the repeater $n'$ and $n_\text{R}(t) \sim \mathcal{CN}(0,\sigma_{\text{R}}^2)$ is the noise of the repeater. The signal retransmitted at repeater $n$, $y_n(t)\in \mathbb{C}$, is 
\begin{equation}
\begin{split}
    \label{eqn:repeater_tx}
    y_n(t) &=\alpha_n u_n(t), \\
    \end{split}
\end{equation} 
where $\alpha_n $$\in$$ \mathbb{C}$ is the amplification gain at repeater $n$. Equation (\ref{eqn:repeater_tx}) can be rewritten in vector considering all $N$ repeaters, $\mathbf{y}_\text{R}(t)\in \mathbb{C}^N$, as 
\begin{equation}
\begin{split}
    \label{eqn:rx_rep_vec}
    \mathbf{y}_\text{R}(t) = \mathbf{\Phi}_\alpha \Bigl( \mathbf{H}_{\text{AR}}^{\top}\mathbf{x}(t) + \mathbf{H}_{\text{RR}}\mathbf{y}_\text{R}(t) + \mathbf{H}_{\text{ADR}}^{\top}\mathbf{x}(t) + \mathbf{n}_\text{R}(t) \Bigl),
\end{split}
\end{equation}
where $\mathbf{H}_{\text{AR}} = [\mathbf{h}_{\text{A},1},\cdots, \mathbf{h}_{\text{A},N}]$ with $\mathbf{H}_{\text{AR}} $$\in$$ \mathbb{C}^{M\times N}$, $\mathbf{\Phi}_\alpha = \diag{[\alpha_1,\dots ,\alpha_N]}$, $\mathbf{H}_{\text{ADR}}=[\mathbf{h}_{\text{AD}
,1},\dots,\mathbf{h}_{\text{AD},N}]$ with $\mathbf{H}_{\text{ADR}} \in \mathbb{C}^{M\times N}$, and $\mathbf{n}_\text{R}(t)\sim \mathcal{CN}(0,\sigma^2_{\text{R}}\mathbf{I}_N)$ is noise at the repeaters. Equation (\ref{eqn:rx_rep_vec}) can be rewritten as 
\[
    \label{eqn:rx_rep_vec_2}
    \mathbf{y}_\text{R}(t) = (\mathbf{I}_N-\mathbf{\Phi}_\alpha\mathbf{H}_{\text{RR}})^{-1}\mathbf{\Phi}_\alpha( \mathbf{H}_{\text{AR}}^{\top}\mathbf{x}(t) + \mathbf{H}_{\text{ADR}}^{\top}\mathbf{x}(t) + \mathbf{n}_\text{R}(t) ).
\]
\subsubsection{Received signal at AP}
We consider the sampled signal received at the AP with symbol index $q$, $\mathbf{y}_{\text{AP}}[q] \in \mathbb{C}^M$ , as
\begin{equation}
\begin{split}
    \label{eqn:rx_ap}
    \mathbf{y}_{\text{AP}}[q] = \underbrace{\mathbf{H}_\text{AD}\mathbf{x}[q]}_\text{Echoes (drone)} + \underbrace{\mathbf{H}_{\text{AR}}\mathbf{y}_\text{R}[q]}_\text{Repeaters} + \underbrace{\mathbf{n}_{\text{AP}}[q]}_\text{Noise (AP)} \\
    = \underbrace{\mathbf{H}_\text{AD}\mathbf{x}[q] + \mathbf{H}_{\text{AR}} (\mathbf{I}_N-\mathbf{\Phi}_\alpha\mathbf{H}_{\text{RR}})^{-1}\mathbf{\Phi}_\alpha \mathbf{H}_{\text{ADR}}^{\top}\mathbf{x}[q] }_\text{Useful signal for sensing} \\
    + \underbrace{\mathbf{H}_{\text{AR}} (\mathbf{I}_N-\mathbf{\Phi}_\alpha\mathbf{H}_{\text{RR}})^{-1}\mathbf{\Phi}_\alpha (\mathbf{H}_{\text{AR}}^{\top}\mathbf{x}[q] + \mathbf{n}_\text{R}[q] ) + \mathbf{n}_{\text{AP}}[q]}_\text{Sensing interference + noise},
\end{split}
\end{equation}
where $\mathbf{n}_{\text{AP}}[q]\sim \mathcal{CN}(0,\sigma^2_{\text{AP}}\mathbf{I}_M)$ is the noise at the AP. 
\subsubsection{Received signal at UE}
The sampled signal received at the UE with symbol index $q$, $y_{\text{UE}}[q] \in \mathbb{C}$, is 
\begin{equation}
\begin{split}
    \label{eqn:UE_rx}
    y_{\text{UE}}[q] &= \mathbf{h}^{\top}_{\text{AU}}\mathbf{x}[q] + n_{\text{UE}}[q], \\
\end{split}
\end{equation}
where $n_{\text{UE}}[q]\sim \mathcal{CN}(0,\sigma_{\text{UE}}^2)$.

\subsection{Precoder}
The precoders in the transmitted signal in (\ref{eqn:tx_AP}) consist of the communication precoder $\mathbf{w}_c$ and the sensing precoder $\mathbf{w}_s$. Given the system model in Fig.\ref{fig:system_model}, note that the link between the AP and repeaters can be masked to avoid the sensing interference term in (\ref{eqn:rx_ap}). To this purpose, we select the normalized communication precoder, $\mathbf{w}_\text{c} = \mathbf{w}'_\text{c}/\norm{\mathbf{w}'_\text{c}}$, where $\mathbf{w}'_\text{c}$ is chosen as the conjugate of the estimated user channel projected onto the orthogonal complement of the subspace spanned by the AP–repeater link as
\begin{equation}
    \mathbf{w}'_\text{c}= (\mathbf{I}_M-\mathbf{a}_\text{n}(0)\mathbf{a}_\text{n}^\mathrm{H}(0))\hat{\mathbf{h}}^*_{\text{AU}},
\end{equation}
where $\mathbf{a}_\text{n}(0)=\mathbf{a}(0)/\norm{\mathbf{a}(0)}$ is the normalized steering vector of the AP–repeater link. To avoid degrading the SINR of the communication link, we select the normalized sensing precoder, $\mathbf{w}_\text{s} = \mathbf{w}'_\text{s}/\norm{\mathbf{w}'_\text{s}}$, where $\mathbf{w}'_\text{s}$ is chosen as the conjugate steering vector in estimated target direction projected onto the orthogonal complement of the subspace spanned by the communication channel and AP-repeater link as
\begin{equation}
\label{eqn:ws}
    \mathbf{w}'_\text{s}=(\mathbf{I}_M-\mathbf{U}(\mathbf{U}^\mathrm{H}\mathbf{U})^{-1}\mathbf{U}^\mathrm{H}) \mathbf{a}^*(\hat{\theta}),
\end{equation}
where columns of $\mathbf{U}$ consist of the normalized steering vector of the AP--repeater link and the normalized user channel estimate, $\mathbf{U=[\mathbf{a}_\text{n}(0), \ \hat{\mathbf{h}}^\ast_{\text{AU}}/\norm{\hat{\mathbf{h}}_{\text{AU}}} ]}$.

\subsection{User SINR}
We consider the case where the AP obtains perfect channel estimates i.e., $\hat{\mathbf{h}}_{\text{AU}}=\mathbf{h}_{\text{AU}}$, hence the user SINR at UE is
\begin{equation}
\begin{split}
    \label{eqn:sinr_ue}
    \gamma_{\text{UE}} &= \frac{\mathbb{E} \{ \abs{ \sqrt{\rho_\text{c}} \mathbf{h}^{\top}_{\text{AU}}  \mathbf{w}_\text{c} }^2 \} }{ \mathbb{E} \{ \abs{ \sqrt{\rho_\text{s}} \mathbf{h}^{\top}_{\text{AU}} \mathbf{w}_\text{s}  + n_{\text{UE}}}^2 \} } \\
    &\approx \frac{ \rho_\text{c} M \beta_{\text{AU}} }{ \sigma^2_{\text{UE}}},
\end{split}
\end{equation}
where $\mathbf{h}^{\top}_{\text{AU}} \mathbf{w}_\text{s}=0$, since $\mathbf{h}^{\top}_{\text{AU}}$ and $\mathbf{w}_\text{s}$ are orthogonal by the precoder design in (\ref{eqn:ws}).
\subsection{Average sensing SINR}
The average sensing SINR, which considers both direct drone echoes and amplified drone echoes as the sensing signal component, treats all remaining terms as interference and noise. The average sensing SINR, $\gamma_s$, can be written as
\begin{equation}
    \label{eqn:sinr_sensing}
    \gamma_\text{s} = \frac{\mathbb{E} \{ \norm{\mathbf{H}_\text{AD}\mathbf{x}+\mathbf{H}_{\text{AR}}(\mathbf{I}_N-\mathbf{\Phi}_\alpha\mathbf{H}_{\text{RR}})^{-1}\mathbf{\Phi}_\alpha \mathbf{H}_{\text{ADR}}^{\top}\mathbf{x} }^2 \} }{ \mathbb{E} \{ \norm{ \mathbf{H}_{\text{AR}}(\mathbf{I}_N-\mathbf{\Phi}_\alpha\mathbf{H}_{\text{RR}})^{-1}\mathbf{\Phi}_\alpha (\mathbf{H}_{\text{AR}}^{\top}\mathbf{x} + \mathbf{n}_\text{R}) + \mathbf{n}_{\text{AP}}} ^2\} }.
\end{equation}
With a large enough repeater spacing $d$, the repeater interaction term, $(\mathbf{I}_N-\mathbf{\Phi}_\alpha\mathbf{H}_{\text{RR}})^{-1}$, can be approximated with first two terms of Neumann series as $(\mathbf{I}_N+\mathbf{\Phi}_\alpha\mathbf{H}_{\text{RR}})$ hence (\ref{eqn:sinr_sensing}) is rewritten as
\begin{equation}
    \label{eqn:sensing_sinr2}
    \gamma_\text{s} = \frac{\mathbb{E} \{ \norm{\mathbf{H}_\text{AD}\mathbf{x}+\mathbf{H}_{\text{AR}}(\mathbf{I}_N+\mathbf{\Phi}_\alpha\mathbf{H}_{\text{RR}})\mathbf{\Phi}_\alpha \mathbf{H}_{\text{ADR}}^{\top}\mathbf{x} }^2 \} }{ \mathbb{E} \{ \norm{ \mathbf{H}_{\text{AR}}(\mathbf{I}_N+\mathbf{\Phi}_\alpha\mathbf{H}_{\text{RR}})\mathbf{\Phi}_\alpha \mathbf{n}_\text{R} + \mathbf{n}_{\text{AP}}} ^2\} },
\end{equation}
where the term $\mathbf{H}_{\text{AR}}^{\top}\mathbf{x}=0$ due to the orthogonality design in (\ref{eqn:ws}).


\section{Repeater gain optimization}
This section will discuss the optimization problem of sensing SINR with respect to the repeater amplification gain.
\subsection{Maximum Amplification Gain}
It is important to constrain the maximum amplification gain, $\alpha_{\text{max}}$, to ensure the stability of the system that is induced by repeater interaction as investigated in \cite[Eq.~8]{10694183}, the $\alpha_{\text{max}}$ is lower bounded by
\[\abs{\alpha_{\text{max}}} \geq \inf_\omega \min_n \frac{1}{\sum_{n'\neq n} \abs{ h_{nn'}(j\omega)}},\]
where $\omega$ is the angular frequency. As we are considering $N$ repeaters arrange in a line with LOS channel between the repeaters, hence the $\alpha_{\text{max}}$ can be selected as 
\begin{equation}
\label{eqn:alpha_max_sta}
    \abs{\alpha_{\text{max}}}= \min_n \frac{1}{\sum_{n'\neq n} \abs{ h_{nn'}(j\omega)}} = \frac{4\pi}{\lambda} \biggl( \max_n \sum_{n'\neq n} \frac{1}{l_{nn'}} \biggl)^{-1}.
\end{equation}
\subsection{Closed-Form Sensing SINR}
The numerator in (\ref{eqn:sensing_sinr2}) can be written as 
\begin{equation}
\label{eqn:numer}
    \mathbb{E} \{ \norm{\mathbf{H}_\text{AD}\mathbf{x} } \} + \mathbb{E} \{ \norm{ \mathbf{H}_{\text{AR}}(\mathbf{I}_N+\mathbf{\Phi}_\alpha\mathbf{H}_{\text{RR}})\mathbf{\Phi}_\alpha \mathbf{H}_{\text{ADR}}^{\top}\mathbf{x} }^2 \},
\end{equation}
where the cross term is zero. The first term in (\ref{eqn:numer}) is
\begin{align*}
    \mathbb{E} \{ \norm{ \mathbf{H}_{\text{AD}} \mathbf{x} }^2\} &= \mathbb{E} \{ \norm{ r \sqrt{\beta_{\text{AD}}} \mathbf{a}(\theta) \mathbf{a}^\top(\theta) \mathbf{x} }^2 \} \\
    &= M \beta_{\text{AD}} \sigma^2_{rcs} \mathbb{E} \{ |\mathbf{a}^\top(\theta) \mathbf{x}|^2 \}\\
    &\approx \sigma^2_{rcs} \beta_{\text{AD}} M\left( \rho_c  + \rho_s M \right).
\end{align*}
The second term in (\ref{eqn:numer}) is
\begin{align*}
    \mathbb{E} \{ \norm{ \mathbf{H}_{\text{AR}} \mathbf{B} \mathbf{H}_{\text{ADR}}^\top \mathbf{x} }^2 \} = \mathbb{E} \{ \norm{ r \mathbf{a}(0) \left( \mathbf{d}^\top \mathbf{B} \mathbf{v} \right) \left( \mathbf{a}^\top(\theta) \mathbf{x} \right) }^2 \},
\end{align*}
where $\mathbf{B} = (\mathbf{I}_N + \mathbf{\Phi}_\alpha \mathbf{H}_{\text{RR}}) \mathbf{\Phi}_\alpha$, $\mathbf{H}_{\text{AR}} = \mathbf{a}(0) \mathbf{d}^\top$ as the repeater positions are specified in integer meters, $\mathbf{H}_{\text{ADR}}^\top = \mathbf{v} \mathbf{a}^\top(\theta)$, $\mathbf{d}=[\sqrt{\beta_{\text{A},1}},\cdots,\sqrt{\beta_{\text{A},N}}]^\top$, and $\mathbf{v}=[\sqrt{\beta_{\text{AD},n}}e^{-j2\pi f_c \tau_{\text{AD},1}},\cdots,\sqrt{\beta_{\text{AD},n}}e^{-j2\pi f_c \tau_{\text{AD},N}}]^\top$. $\xi(\boldsymbol{\alpha})= \mathbf{d}^\top \mathbf{B} \mathbf{v} $ denotes the effective repeater gain and is equal to
\begin{equation}
\begin{split}
    \xi(\boldsymbol{\alpha}) = \sum_{n=1}^N \alpha_n \sqrt{\beta_{\text{A},n}\beta_{\text{AD},n}} e^{-j2\pi f_c \tau_{\text{AD},n}} \\
    + \frac{\lambda}{4\pi }\sum_{n=1}^N \sum_{m \neq n} \alpha_n \alpha_m \frac{1}{\abs{m-n}d} \sqrt{\beta_{\text{A},n} \beta_{\text{AD},m}} e^{-j2\pi f_c \tau_{\text{AD},m}},
\end{split}
\end{equation}
where $\boldsymbol{\alpha}=[\alpha_1,\alpha_2,\cdots,\alpha_N]^\top$. The term $\mathbb{E} \{ |\mathbf{a}^\top(\theta) \mathbf{x}|^2 \}$ is
\begin{align*}
    \mathbb{E} \{ |\mathbf{a}^\top(\theta) \mathbf{x}|^2 \} &= \rho_c |\mathbf{a}^\top(\theta) \mathbf{x}_c|^2 + \rho_s |\mathbf{a}^\top(\theta) \mathbf{x}_s|^2 \\
    &\approx \rho_c + \rho_sM,
\end{align*}
and therefore,
\begin{align*}
     \mathbb{E} \{ \norm{ \mathbf{H}_{\text{AR}} \mathbf{B} \mathbf{H}_{\text{ADR}}^\top \mathbf{x} }^2 \} \approx M \sigma_{\text{rcs}}^2 |\xi(\boldsymbol{\alpha})|^2 (\rho_c + \rho_s M).
\end{align*}
Combining both terms, (\ref{eqn:numer}) is approximated as 
\begin{equation}
    Mc_1 + M c_2 |\xi(\boldsymbol{\alpha})|^2,
\end{equation}
where $c_1=\sigma^2_{rcs} \beta_{\text{AD}} \left( \rho_c  + \rho_s M \right)$ and $c_2=\sigma^2_{rcs}(\rho_c + \rho_s M)$. The denominator in (\ref{eqn:sensing_sinr2}) is rewritten as
\begin{equation}
\label{eqn:denom}
    \mathbb{E} \{ \lVert \mathbf{H}_{\text{AR}} \mathbf{B} \mathbf{n}_{\text{R}} \rVert^2 \} + \mathbb{E} \{ \lVert \mathbf{n}_{\text{AP}} \rVert^2 \},
\end{equation}
where the cross term is zero, again $\mathbf{B}= (\mathbf{I}_N + \mathbf{\Phi}_\alpha \mathbf{H}_{\text{RR}}) \mathbf{\Phi}_\alpha$. The second term in (\ref{eqn:denom}) is $\mathbb{E} \{ \lVert \mathbf{n}_{\text{AP}} \rVert^2 \} = M \sigma_{\text{AP}}^2$, and the first term in (\ref{eqn:denom}) is 
\begin{align*}
    \mathbb{E} \{ \lVert \mathbf{H}_{\text{AR}} \mathbf{B} \mathbf{n}_{\text{R}} \rVert^2 \} &= \sigma_{\text{R}}^2 \text{Tr}(\mathbf{H}_{\text{AR}} \mathbf{B} \mathbf{B}^\mathrm{H} \mathbf{H}_{\text{AR}}^\mathrm{H}) \\
    &= \sigma_{\text{R}}^2 \text{Tr}(\mathbf{H}_{\text{AR}}^\mathrm{H} \mathbf{H}_{\text{AR}} \mathbf{B} \mathbf{B}^\mathrm{H}) \\
    &= \sigma_{\text{R}}^2 M \cdot \text{Tr}( \mathbf{d} \mathbf{d}^\top \mathbf{B} \mathbf{B}^\mathrm{H}) \\
    &= M \sigma_{\text{R}}^2 \lVert \mathbf{B}^\mathrm{H} \mathbf{d} \rVert^2.
\end{align*} 
The denominator in (\ref{eqn:sensing_sinr2}) is
\begin{equation}
    M \left( \sigma_{\text{R}}^2 \lVert \mathbf{B}^\mathrm{H} \mathbf{d} \rVert^2 + \sigma_{\text{AP}}^2 \right),
\end{equation}
and therefore the $\gamma_s$ in (\ref{eqn:sensing_sinr2}) is
\begin{equation}
    \gamma_s = \frac{c_1 + c_2 \abs{ \mathbf{d}^\top \mathbf{B} \mathbf{v} }^2 }{ \sigma_{\text{R}}^2 \lVert \mathbf{B}^\mathrm{H} \mathbf{d} \rVert^2 + \sigma_{\text{AP}}^2}.
\end{equation}
\subsection{Amplification Gain Optimization}
We consider a sensing-centric optimization while satisfying communication requirements. Thus, we formulate the maximization of sensing SINR as (\ref{eqn:sensing_sinr2}) with respect to $\mathbf{\Phi_\alpha}$. The optimization problem is formulated as
\begin{subequations}
\label{eqn:opt1}
\begin{align}
\max_{\boldsymbol{\alpha}}  \quad
& \gamma_\text{s}(\boldsymbol{\alpha}) = \frac{c_1 + c_2 \abs{\mathbf{d}^\top \mathbf{B} \mathbf{v}}^2}{ \sigma_{\text{R}}^2  (\mathbf{d}^\top \mathbf{B} \mathbf{B}^\mathrm{H} \mathbf{d})  + \sigma_{\text{AP}}^2}\\
\text{s.t.}\quad 
& 0 \le \abs{\alpha_n} \le \abs{\alpha_{\max}}, \ \forall n, \label{c:alpha} \\
& \gamma_{\text{UE}} = \frac{ \rho_\text{c} M \beta_{\text{AU}} }{\sigma_{\text{UE}}^2} \ \ge\ \gamma_{\text{UE,req}}, \label{c:UE}\\
& \rho_\text{c} + \rho_\text{s} \ \le\ \rho_{\max} \label{c:power},
\end{align}
\end{subequations}
where $\alpha_n \in \mathbb{C}$ is the amplification gain at repeater $n$, $\rho_{\text{max}}$ is maximum transmitted power at the AP, and $\gamma_{\text{UE,req}}$ is user SINR requirement for communication. Note that the signal power can be solved separately as $\rho_c=\frac{\gamma_{\text{UE,req}}\sigma_{\text{UE}}^2}{\rho_cM}$, and $\rho_s=\rho_{\text{max}}-\rho_c$. For simplicity, we assume that the positive-feedback effect in the sensing SINR is negligible in the optimization problem, i.e., the repeaters are sufficiently far apart, although this effect is taken into account in the simulations. Consequently, the coupling matrix simplifies to $\mathbf{B} = \mathbf{\Phi}_\alpha$. The numerator term $\mathbf{d}^\top \mathbf{B} \mathbf{v}$ is 
\begin{equation}
\mathbf{d}^\top \mathbf{B} \mathbf{v} = \mathbf{d}^\top \mathbf{\Phi}_\alpha \mathbf{v} = \sum_{n=1}^N d_n v_n \alpha_n = \sum_{n=1}^N a_n \alpha_n 
\end{equation}
where $a_n = d_n v_n$ and the denominator term $\mathbf{d}^\top \mathbf{B} \mathbf{B}^\mathrm{H} \mathbf{d} = \sum_{n=1}^N d_n^2 \vert{}\alpha_n\vert{}^2$.
The simplified optimization problem is
\begin{equation}
    \begin{split}
    \label{eq:simplified_opt}
   \max_{\boldsymbol{\alpha}} \quad \gamma_\text{s}(\boldsymbol{\alpha}) = \frac{c_1 + c_2 \left\vert{} \sum_{n=1}^N a_n \alpha_n \right\vert{}^2}{\sigma_{\text{R}}^2 \sum_{n=1}^N d_n^2 \vert{}\alpha_n\vert{}^2 + \sigma_{\text{AP}}^2} \\
   \text{s.t.} \quad 0 \le \vert{}\alpha_n\vert{} \le \vert{}\alpha_{\max}\vert{}, \quad \forall n = 1, \dots, N. 
   \end{split}
\end{equation}
Let $\alpha_{\text{sol},n} = r_n e^{j \phi_n}$ be the sub-optimal solution for the $n$-th repeater obtained from the optimization problem \eqref{eq:simplified_opt}. The sub-optimal phase choice is 
\begin{equation}
\phi_n = -\arg(a_n) = -\arg(d_n v_n).
\end{equation}
The sub-optimal magnitude is
\begin{equation}
r_n = \min \left( \mu \frac{\vert{}v_n\vert{}}{\vert{}d_n\vert{}}, \ \vert{}\alpha_{\max}\vert{} \right),
\end{equation}
where $\mu > 0$ is a scaling parameter.

\section{Simulation results}
\begin{small}
\begin{table}
    \caption{Baseline parameters.}
    \vspace{-0.4cm}
    \begin{center}
        \begin{tabular}{c|c|c|c   }
            \hline
            \textbf{Parameter}  & \textbf{Value} & \textbf{Parameter}  & \textbf{Value} \\
            \hline
            $M$  & $100$  &      $\sigma_\text{RCS}^2$ & $-10$ dBsm \\
            \hline
            $f_\text{c}$ & $15$ GHz & $\sigma_\text{R}^2$ & $-110$ dBm   \\
            \hline
            $l_\text{AD}$ & $500$ m  & $\sigma_{\text{AP}}^2,\sigma_{\text{UE}}^2$ & $-90$ dBm
            \\ 
            \hline
            $l_{\text{AU}}$ & $200$ m  &  $d$ & $100$ m   \\
            \hline
            $l_{\text{A},1}$ & $250$ m & $\gamma_{\text{UE,req}}$ & $30$ dB \\
            \hline
            $\theta$ & $\pi/6$ \\
            \hline
            
        \end{tabular}
        \label{tab:params}
    \end{center}
    \label{table:simparams}
\end{table}
\end{small}
In this section, we investigate the sensing performance by using the sensing SINR as a performance metric. The setup follows the system model in Fig.~\ref{fig:system_model}, where the distance between the AP and the drone is $l_{\text{AD}}=500$ m. The repeater closest to the AP is the first repeater, located $l_{\text{A},1}=250$ m away from the AP at the same altitude. The repeaters are arranged in a line with a uniform spacing of $d=100$ m, which result in $\abs{\alpha_{\text{max}}}$$=$$41.6$ dB as determined in~\eqref{eqn:alpha_max_sta}. The baseline parameters are shown in Table \ref{tab:params}, unless explicitly stated otherwise. 
\begin{figure}[htbp] 
    \centering
        \resizebox{1.0\linewidth}{!}{
%
%
\begin{tikzpicture}

\begin{axis}[%
width=6.028in,
height=4.754in,
at={(1.011in,0.642in)},
scale only axis,
xmin=52,
xmax=53,
xlabel style={font=\huge\color{white!15!black}},
xlabel={$\rho_{\textrm{max}}$ [dBm]},
ymode=normal,
ymin=0.00501849220449745,
ymax=0.00894196404317179,
yminorticks=true,
ylabel style={font=\huge\color{white!15!black}},
ylabel={$\gamma_s$},
axis background/.style={fill=white},
xmajorgrids,
ymajorgrids,
yminorgrids,
legend cell align={left},
legend style={font=\LARGE, at={(0.05,0.95)}, anchor=north west, draw=white!15!black}
]
\addplot [color=black, line width=4.0pt, mark=o, mark size=4pt]
  table[row sep=crcr]{%
52	0.00709418332877539\\
52.2	0.00744095988248857\\
52.4	0.0077680062074303\\
52.6	0.00814876807000419\\
52.8	0.00855176758507727\\
53	0.00894196404317179\\
};
\addlegendentry{${\alpha}=\alpha_{\textrm{sol}}, N=5$}

\addplot [color=black, line width=4.0pt, mark=+, mark size=5pt]
  table[row sep=crcr]{%
52	0.00557900056618376\\
52.2	0.00588078621876341\\
52.4	0.0062204594401215\\
52.6	0.00643383513048756\\
52.8	0.00674455661225338\\
53	0.00709458971745044\\
};
\addlegendentry{$\alpha=\alpha_{\textrm{max}}, N=5$}

\addplot [color=black, dotted, line width=4.0pt, mark=diamond, mark size=4pt]
  table[row sep=crcr]{%
52	0.00501849220449745\\
52.2	0.0052594308937148\\
52.4	0.00550206630971799\\
52.6	0.00583641407660272\\
52.8	0.00600289466422676\\
53	0.0063437648780251\\
};
\addlegendentry{without repeaters}

\addplot [color=black, dashed, line width=4.0pt, mark=triangle, mark size=4pt]
  table[row sep=crcr]{%
52	0.00581966554372477\\
52.2	0.00611167273196002\\
52.4	0.00638104596713714\\
52.6	0.00663381465221663\\
52.8	0.00704079387149768\\
53	0.00732002424095661\\
};
\addlegendentry{${\alpha}=\alpha_{\textrm{sol}}, N=20$}

\end{axis}

\begin{axis}[%
width=7.778in,
height=5.833in,
at={(0in,0in)},
scale only axis,
xmin=0,
xmax=1,
ymin=0,
ymax=1,
axis line style={draw=none},
ticks=none,
axis x line*=bottom,
axis y line*=left
]
\end{axis}
\end{tikzpicture}%
        } 
        \caption{Average sensing SINR $\gamma_s$.}
    \label{fig:sensing_SINR}
\end{figure}
\begin{figure}[htbp] 
    \centering
        \resizebox{1.0\linewidth}{!}{
%
%
\begin{tikzpicture}

\begin{axis}[%
width=6.028in,
height=4.754in,
at={(1.011in,0.642in)},
scale only axis,
xmin=52,
xmax=53,
xlabel style={font=\huge\color{white!15!black}},
xlabel={$\rho_{\textrm{max}}$ [dBm]},
ymode=normal,
ymin=0.00506278976882439,
ymax=0.0144979534342683,
yminorticks=true,
ylabel style={font=\huge\color{white!15!black}},
ylabel={$\gamma_s$},
axis background/.style={fill=white},
xmajorgrids,
ymajorgrids,
yminorgrids,
legend style={font=\LARGE, at={(0.05,0.97)}, anchor=north west, legend cell align=left, align=left, draw=white!15!black}
]
\addplot [color=black, dashed, line width=4.0pt, mark=+, mark size=5pt]
  table[row sep=crcr]{%
52	0.0114114425672745\\
52.2	0.0118765220532715\\
52.4	0.0125101943082191\\
52.6	0.0130592265322648\\
52.8	0.0137331391551056\\
53	0.0144979534342683\\
};
\addlegendentry{$N\alpha_{\hat{n},\textrm{sol}}, \textrm{1-active}$}

\addplot [color=black, line width=4.0pt, mark=square, mark size=5pt]
  table[row sep=crcr]{%
52	0.00583910282206937\\
52.2	0.00612976707079297\\
52.4	0.00644265991591543\\
52.6	0.006728983480603\\
52.8	0.00697464529551022\\
53	0.00734910458999774\\
};
\addlegendentry{${\alpha}=\alpha_{\textrm{sol}}, N=20$}

\addplot [color=black, dotted, line width=4.0pt, mark=diamond, mark size=5pt]
  table[row sep=crcr]{%
52	0.00506278976882439\\
52.2	0.0053199296105178\\
52.4	0.00556301971755395\\
52.6	0.00579627997239888\\
52.8	0.00606416136931508\\
53	0.00635716697364167\\
};
\addlegendentry{without repeaters}

\end{axis}

\begin{axis}[%
width=7.778in,
height=5.833in,
at={(0in,0in)},
scale only axis,
xmin=0,
xmax=1,
ymin=0,
ymax=1,
axis line style={draw=none},
ticks=none,
axis x line*=bottom,
axis y line*=left
]
\end{axis}
\end{tikzpicture}%
        } 
        \caption{Proposed method (1-active).}
    \label{fig:proposed}
\end{figure}

Fig. \ref{fig:sensing_SINR} shows the average sensing SINR $\gamma_s$ as a function of $\rho_{\text{max}}$ for different repeater gains. We observe that the optimized gain $\bm\alpha=\bm\alpha_{\text{sol}}$ outperforms marginally both the case where all repeaters are set to maximum gain $\bm\alpha=\bm\alpha_{\text{max}}$ and the case without repeaters $\bm\alpha=0$. Note that, for $\bm\alpha_{\text{sol}}$, there is a dominant gain (i.e., a gain with a large complex magnitude) for the repeater located approximately closest to the drone, while the gains for the surrounding repeaters near that particular repeater are relatively small. Note that the setup involves only a single drone, corresponding to a single dominant echo direction. We further investigate a scenario in which the number of repeaters is increased to $N=20$. The results indicate that the average sensing performance with $N=20$ is worse than that achieved with $N=5$. This behavior can be attributed to the stronger positive feedback induced by the larger number of repeaters. In addition, the constraint on $\abs{\bm\alpha_{\text{max}}}$ in (\ref{eqn:opt1}b) becomes increasingly restrictive because $\abs{\bm\alpha_{\text{max}}}$ decreases, since the repeater gains are non-zero for all $N$.

As the noise‑echo effect from repeater interactions is observed in Fig.~\ref{fig:sensing_SINR}, we propose a basic algorithm that selects a single active repeater, $\hat{n}$, corresponding to the index of the dominant gain from the optimizer, i.e.,
\[\hat{n}=\argmax_n \;\abs{\alpha_{\text{sol},n}},\]
and all remaining repeaters are set to an inactive state. In addition, we boost the gain of the dominant repeater by a factor of $N$ as $\alpha_{\hat{n}}=N\alpha_{\hat{n},\text{sol}}$, while the remaining repeaters remain inactive with $\alpha_n=0$ for $n\neq\hat{n}$ ensuring a fair comparison with the case of $N$ active repeaters. Note that the maximum amplification‑gain constraint $\alpha_{\text{max}}$ is no longer relevant, since only a single repeater is active. The result in Fig.~\ref{fig:proposed} shows that the proposed method outperforms the case where all repeaters are active with the optimized gain $\bm\alpha_{\text{sol}}$.

Although the results indicate that a large number of active repeaters is not favorable, the repeater-swarm architecture is still useful because some repeaters must be located close to the unknown target position to achieve better $\gamma_s$. Most importantly, the system should operate dynamically so that the majority of repeaters remain in the off-state, while only single, or a small subset, becomes active when a target is present.


\vspace{-0.3cm}
\section{Conclusion}
\vspace{-0.2cm}
This paper investigates drone detection in a swarm repeater-assisted MIMO ISAC system. A monostatic setup is considered, with a known target angle and a LOS channel to the drone. The AP simultaneously serves a user in a dense urban area and senses a target in a weak coverage region. The amplification gain of the repeaters is optimized to maximize the average sensing SINR. The results show that the optimized solution assigns one dominant gain to the repeater closest to the drone, while neighboring repeaters receive small gains. Although having a large number of active repeaters within a cluster is undesirable due to positive feedback, the repeater‑swarm architecture remains valuable because some repeaters must be near the target to achieve good sensing SINR. Therefore, the paper proposes a single active‑repeater selection algorithm, which outperforms the cluster‑repeater approach.

\section{Funding Acknowledgments}
This publication has been supported by the Swedish strategic research environment ELLIIT and the WASP-funded project ``ALERT''.

\bibliographystyle{IEEEtran}
\bibliography{refs}

\end{document}